\documentclass[lettersize,journal]{IEEEtran}
\usepackage{amsmath,amsfonts, amssymb}
\usepackage{algorithmic}
\usepackage{algorithm}
\usepackage{array}
\usepackage[caption=false,font=normalsize,labelfont=sf,textfont=sf]{subfig}
\usepackage{textcomp}
\usepackage{tikz}
\usepackage{tkz-euclide}
\usepackage{stfloats}
\usepackage{url}
\usepackage{verbatim}
\usepackage{numprint}
\usepackage{enumitem}
\usepackage{graphicx}
\usepackage{cite}
\usepackage{svg}
\def\clap#1{\hbox to 0pt{\hss#1\hss}}

\begin{document}

\title{Multi-view Disparity Estimation Using\\a Novel Gradient Consistency Model}

\author{James L. Gray, \IEEEmembership{Student Member, IEEE}, Aous T. Naman, \IEEEmembership{Senior Member, IEEE}, David S. Taubman, \IEEEmembership{Fellow, IEEE}
\thanks{The authors are with the School of Electrical Engineering and Telecommunications, University of New South Wales, Sydney, Australia.}
\thanks{This research includes computations using the computational cluster Katana supported by Research Technology Services at UNSW Sydney.}
\thanks{}
}

\markboth{IEEE Transactions on Image Processing,~Vol.~XX, No.~X, Month~2024}%
{Gray \MakeLowercase{\textit{et al.}}: Multi-view Disparity Estimation}

\IEEEpubid{This work has been submitted to the IEEE for possible publication. Copyright may be transferred without notice, after which this version may no longer be accessible.}

\maketitle

\begin{abstract}
Variational approaches to disparity estimation typically use a linearised brightness constancy constraint, which only applies in smooth regions and over small distances. Accordingly, current variational approaches rely on a schedule to progressively include image data. 
This paper proposes the use of Gradient Consistency information to assess the validity of the linearisation; this information is used to determine the weights applied to the data term as part of an analytically inspired Gradient Consistency Model. 
The Gradient Consistency Model penalises the data term for view pairs that have a mismatch between the spatial gradients in the source view and the spatial gradients in the target view.
Instead of relying on a tuned or learned schedule, the Gradient Consistency Model is self-scheduling, since the weights evolve as the algorithm progresses. 
We show that the Gradient Consistency Model outperforms standard coarse-to-fine schemes and the recently proposed progressive inclusion of views approach in both rate of convergence and accuracy. 
\end{abstract}


\section{Introduction}
Scene flow, optical flow and disparity estimation are closely related fields of research that have centred on estimating displacements between corresponding points in images of the same scene. 
These images may be taken at different instances in time, as per optical flow, taken at different locations in space as per disparity estimation or both in the case of the scene flow. 
These tasks have clear applications for higher level computer vision tasks such as autonomous robot navigation and driving, object tracking, action recognition, segmentation, and product quality inspection.

Estimating displacement fields can be cast as an energy minimisation problem with two terms, a data term based on the brightness constancy constraint and a regularisation or smoothing term~\cite{TU2019_Survey, FORTUN20151, Hamzah_Stereo_lit_2015}. 
Strategies for solving the energy minimisation problem can broadly be classed as: variational approaches; search and cost volume based approaches; or hybrid methods.

Search and cost volume based approaches consider multiple candidate displacement fields simultaneously and then select the best candidate. 
To do this, the energy minimisation problem is discretised. 
Markov Random Field frameworks are typically used to treat the energy minimisation task as a labelling problem~\cite{Hamzah_Stereo_lit_2015, FORTUN20151, Chen2016_Full_Flow, Xu_Optical_Flow_Cost_Volume}. 

Many machine learning approaches to scene flow, optical flow and depth estimation incorporate a serach-based approach. 
Some explicitly have cost or correlation volumes based on learned features~\cite{CRAFT2022, RAFT2020, Laga2022DeepSurvery, Huang_2022_FlowFormer}. 
Some approaches may not have an explicit cost or correlation volume, such as PerceiverIO~\cite{Jaegle2021PerceiverIA} or FlowNetSimple~\cite{Dosovitskiy2015}.
However, PerceiverIO  uses \textit{Attention Scores}~\cite{Jaegle2021PerceiverIA} and FlowNetSimple uses many channels in the middle stage of a U-Net architecture~\cite{Dosovitskiy2015}.
In effect, these structures allow multiple candidates to be evaluated simultaneously in accordance with the number of channels.

Hybrid methods include feature based techniques, in which a serach-based approach yields an initial sparse displacement field, which is subsequently interpolated and subjected to variational refinement~\cite{Xu_Motion_Detail_Optical_Flow, EpicFlow_2015, Hu_2016_CVPR}.

Our paper focuses on variational or continuous approaches. 
Variational approaches are arguably simpler than serach-based methods or hybrid methods. This is because variational approaches do not need to compare multiple alternate hypotheses concerning the displacement field and instead only iteratively refine one candidate displacement field until it converges.
However, variational approaches assume that the brightness constancy constraint can be locally linearised~\cite{FORTUN20151, TU2019_Survey, Tran2017Disparity, Young2019, Tran2020_Disparity, Roxas_2020_FisheyeStereo, rao_optical_2023, Ma2019}. 
Whilst such an assumption holds for small displacements and/or low-frequency data, it does not hold for large displacements and high-frequency content~\cite{FORTUN20151}.

The standard approach to address the limited range over which the linearisation is valid has been the use of coarse-to-fine schemes.
A coarse-to-fine scheme initially blurs the image data, so that the linearisation is valid over longer distances~\cite{FORTUN20151, TU2019_Survey, Tran2017Disparity, Tran2020_Disparity, rao_optical_2023}. 
\IEEEpubidadjcol
However, these strategies have a tendency to over-smooth small details at coarse scales, producing errors that cannot be subsequently corrected at finer scales~\cite{EpicFlow_2015, Deng2021_CTF}.

Building on our previous work~\cite{JLGray_OpticalFlow2022}, we present an alternate approach to dealing with the limited valid range of the linearisation.
We propose using Gradient Consistency information to assess the reliability of the linearisation directly; this reliability information is used to derive appropriate weights for the data term at each location in each view and each scale. 

It is important to note that the use of Gradient Consistency information in this work is distinct from a gradient constancy assumption that is commonly used~\cite{brox2004high, Tran2017Disparity, Tran2020_Disparity, rao_optical_2023}.
A gradient constancy assumption, just like the brightness constancy assumption can be a component of the observation model for disparity estimation.
However, Gradient Consistency as used in this paper, refers to the consistency of the equations that arise when linearising that model.

\begin{figure*}[!t]
\centering
\scalebox{0.5}{%
\begin{tikzpicture}[fill=yellow!20]
\usetikzlibrary {shapes.geometric}
\usetikzlibrary {arrows.meta} 
        \draw[left color=white,right color=black!40,middle color=red!40] (2,-0.75,0) -- (2,1.25,0)-- (4,1.25,0) -- (4,-0.75,0) -- cycle;  
        \draw[left color=white,right color=black!40,middle color=red!40] (2,-0.75,0.2) -- (2,1.25,0.2)-- (4,1.25,0.2) -- (4,-0.75,0.2) -- cycle;  
        \draw[left color=white,right color=black!40,middle color=red!40] (2,-0.75,0.4) -- (2,1.25,0.4)-- (4,1.25,0.4) -- (4,-0.75,0.4) -- cycle;  

        \draw[left color=white,right color=black!40,middle color=red!40] (-1,-0.75,0) -- (-1,1.25,0)-- (1,1.25,0) -- (1,-0.75,0) -- cycle;  
        \draw[left color=white,right color=black!40,middle color=red!40] (-1,-0.75,0.2) -- (-1,1.25,0.2)-- (1,1.25,0.2) -- (1,-0.75,0.2) -- cycle;  
        \draw[left color=white,right color=black!40,middle color=red!40] (-1,-0.75,0.4) -- (-1,1.25,0.4)-- (1,1.25,0.4) -- (1,-0.75,0.4) -- cycle;  

        \draw[left color=white,right color=black!40,middle color=red!40] (-4,-0.75,0) -- (-4,1.25,0)-- (-2,1.25,0) -- (-2,-0.75,0) -- cycle;  
        \draw[left color=white,right color=black!40,middle color=red!40] (-4,-0.75,0.2) -- (-4,1.25,0.2)-- (-2,1.25,0.2) -- (-2,-0.75,0.2) -- cycle;  
        \draw[left color=white,right color=black!40,middle color=red!40] (-4,-0.75,0.4) -- (-4,1.25,0.4)-- (-2,1.25,0.4) -- (-2,-0.75,0.4) -- cycle;  


        \draw[left color=white,right color=black!40,middle color=red!40] (2,-3.75,0) -- (2,-1.75,0)-- (4,-1.75,0) -- (4,-3.75,0) -- cycle;  
        \draw[left color=white,right color=black!40,middle color=red!40] (2,-3.75,0.2) -- (2,-1.75,0.2)-- (4,-1.75,0.2) -- (4,-3.75,0.2) -- cycle;  
        \draw[left color=white,right color=black!40,middle color=red!40] (2,-3.75,0.4) -- (2,-1.75,0.4)-- (4,-1.75,0.4) -- (4,-3.75,0.4) -- cycle;  

        \draw[left color=white,right color=black!40,middle color=red!40] (-1,-3.75,0) -- (-1,-1.75,0)-- (1,-1.75,0) -- (1,-3.75,0) -- cycle;  
        \draw[left color=white,right color=black!40,middle color=red!40] (-1,-3.75,0.2) -- (-1,-1.75,0.2)-- (1,-1.75,0.2) -- (1,-3.75,0.2) -- cycle;  
        \draw[left color=white,right color=black!40,middle color=red!40] (-1,-3.75,0.4) -- (-1,-1.75,0.4)-- (1,-1.75,0.4) -- (1,-3.75,0.4) -- cycle;  

        \draw[left color=white,right color=black!40,middle color=red!40] (-4,-3.75,0) -- (-4,-1.75,0)-- (-2,-1.75,0) -- (-2,-3.75,0) -- cycle;  
        \draw[left color=white,right color=black!40,middle color=red!40] (-4,-3.75,0.2) -- (-4,-1.75,0.2)-- (-2,-1.75,0.2) -- (-2,-3.75,0.2) -- cycle;  
        \draw[left color=white,right color=black!40,middle color=red!40] (-4,-3.75,0.4) -- (-4,-1.75,0.4)-- (-2,-1.75,0.4) -- (-2,-3.75,0.4) -- cycle;  


        \draw[left color=white,right color=black!40,middle color=red!40] (2,-6.75,0) -- (2,-4.75,0)-- (4,-4.75,0) -- (4,-6.75,0) -- cycle;  
        \draw[left color=white,right color=black!40,middle color=red!40] (2,-6.75,0.2) -- (2,-4.75,0.2)-- (4,-4.75,0.2) -- (4,-6.75,0.2) -- cycle;  
        \draw[left color=white,right color=black!40,middle color=red!40] (2,-6.75,0.4) -- (2,-4.75,0.4)-- (4,-4.75,0.4) -- (4,-6.75,0.4) -- cycle; 

        \draw[left color=white,right color=black!40,middle color=red!40] (-1,-6.75,0) -- (-1,-4.75,0)-- (1,-4.75,0) -- (1,-6.75,0) -- cycle;  
        \draw[left color=white,right color=black!40,middle color=red!40] (-1,-6.75,0.2) -- (-1,-4.75,0.2)-- (1,-4.75,0.2) -- (1,-6.75,0.2) -- cycle;  
        \draw[left color=white,right color=black!40,middle color=red!40] (-1,-6.75,0.4) -- (-1,-4.75,0.4)-- (1,-4.75,0.4) -- (1,-6.75,0.4) -- cycle; 

        \draw[left color=white,right color=black!40,middle color=red!40] (-4,-6.75,0) -- (-4,-4.75,0)-- (-2,-4.75,0) -- (-2,-6.75,0) -- cycle;  
        \draw[left color=white,right color=black!40,middle color=red!40] (-4,-6.75,0.2) -- (-4,-4.75,0.2)-- (-2,-4.75,0.2) -- (-2,-6.75,0.2) -- cycle;  
        \draw[left color=white,right color=black!40,middle color=red!40] (-4,-6.75,0.4) -- (-4,-4.75,0.4)-- (-2,-4.75,0.4) -- (-2,-6.75,0.4) -- cycle; 

        \draw[dashed] (-4.5,-7.25,0) -- (-4.5,1.5,0)-- (4.25,1.5,0) -- (4.25,-7.25,0) -- cycle;
        

        \path (
            (-4.5,-7.25,0) node(BL_BOX) [] {}
            (-4.5,1.5,0) node(TL_BOX) [] {}
            (4.25,1.5,0) node(TR_BOX) [] {}
            (4.25,-7.25,0) node(BR_BOX) [] {}
            (9, -2.75, 0)  node(d) [rectangle, draw] {}
            (11, -1.5, 0) node(e) [ellipse,draw]            {\Large GC Model}
            (11, -2.75, 0) node(f) [circle, draw] {{\Large$\times$}}
            (4.25, -6, 0)  node(g) [] {}
            (17, -2.75, 0) node(k) [ellipse, draw] {\Large Optimisation}
            (17, 0.75, 0) node(l) [ellipse, draw] {\Large Regulariser}
            (13, -2.75, 0) node(DAT) [above] {\Large Data Term}
            (17, -1.0, 0) node(REG) [above, rotate=-90] {\Large Reg. Term}
            (17, -5, 0) node(TopMiddle) []{}
            (17, -9, 0) node (DIS) [below] {\Large Disparity Field}
            (0, -7.5, 0) node (IM_Label) [below] {\Large Multiple Views at Multiple Scales}
            ;
        \draw[dashed] (TR_BOX) -- (d);
        \draw[dashed] (BR_BOX) -- (d);
        
        \draw[thick, -stealth] (d) -- (e);
        \draw[thick, -stealth] (d) -- (f);
        \draw[thick, -stealth] (e) -- (f);
        \draw[thick, -stealth] (f) -- (k);
        \draw[thick, -stealth] (l) -- (k);
        \draw[left color=yellow,right color=blue,middle color=green!50] (15, -5, 0) -- (19, -5, 0) -- (19, -9, 0) -- (15, -9, 0) -- cycle;
        \draw[thick, -stealth] (k) -- (TopMiddle);

\end{tikzpicture}
}
\caption{In this diagram, we detail how we apply the Gradient Consistency Model (GCM) to a multi-scale context. We use multiple views and multiple scales simultaneously to estimate the disparity field. Each scale and view is weighted using the GCM and then the data is passed to the optimisation stage which determines the disparity field.}
\label{fig:Multi-Scale_Diagram}
\end{figure*}
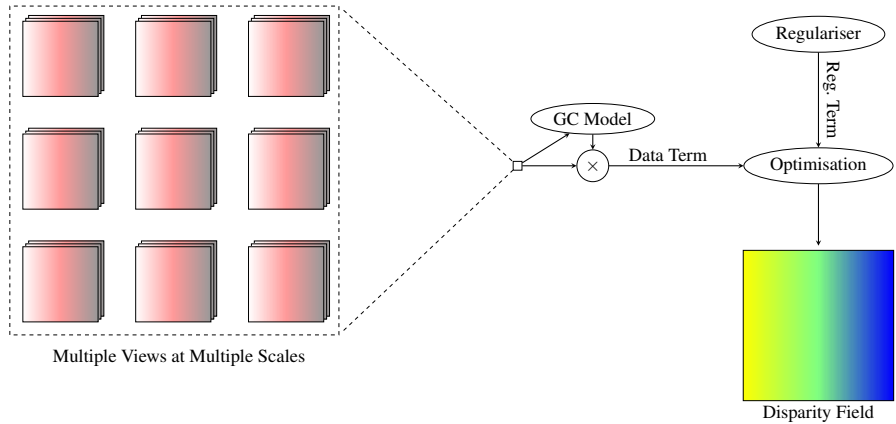

To demonstrate the benefit of using Gradient Consistency information, we develop an analytical Gradient Consistency Model (GCM) and apply it to single-scale and multi-scale multi-view disparity estimation. 
In the multi-scale case, the GCM uses multiple views and multiple scales simultaneously to estimate the disparity field as shown in Figure~\ref{fig:Multi-Scale_Diagram}. 
Each scale and view is weighted using the GCM and then the data is used to determine the displacement field.
Unlike previous approaches such as coarse-to-fine schemes which rely on a manually determined schedule, the GCM does not; 
primarily, the GCM sets its weights based on data-driven estimates of Gradient Consistency.
These weights evolve over the course of the algorithm, as it converges. 
Therefore, the GCM is essentially self-scheduling, relying on only two manually set parameters: an image acquisition noise term; and a regularisation parameter, as found in all variational schemes.
We find that the GCM is substantially insensitive to changes in these parameters and outperforms a coarse-to-fine approach both in terms of rate of convergence and accuracy.

This paper is organised as follows, we introduce Gradient Consistency in Section~\ref{sec:Intro_to_GC}.
Our multi-scale framework and Scale Consistency is introduced in Section~\ref{sec:ScaleConsistency}.
In Section~\ref{sec:Physical_Coupling}, we discuss the physical coupling between the displacement fields of different views in multi-view disparity estimation.
The Gradient Consistency Model is detailed in Section~\ref{sec:GCM}.
Our GCM is applied to a variational approach in Section~\ref{sec:VariationalModel}.
The performance of the GCM in a single scale context is evaluated in Section~\ref{sec:Single_Scale_Results} and the multi-scale results are discussed in Section~\ref{sec:Eval}.
Conclusions are provided in Section~\ref{sec:Conclusion}.

\section{Introduction to Gradient Consistency} \label{sec:Intro_to_GC}
Variational approaches to disparity, scene flow and optical flow estimation use an iterative approach. Each iteration involves two stages: first, we estimate residual displacements to update the estimated absolute displacements; then based on the updated absolute displacements, we warp the images so that they become as similar as possible. These iterations are repeated until the variational scheme converges on the final absolute estimated displacements.

The simplest approaches to estimating residual displacements at each iteration aim to satisfy the brightness constancy constraint \cite{FORTUN20151, TU2019_Survey, Hamzah_Stereo_lit_2015, Ma2019}. 
This process involves solving a series of coupled brightness constancy constraint equations between image pairs. 
We can write the brightness constancy constraint equation between a reference image, $I_{r}$ and a target image $I_{t}$ as,
\begin{equation}\label{eqn:Generalised_Brightness_Constancy}
    I_{r}(\mathbf{s}) - I_{t}(\mathbf{s} + \delta \mathbf{v}_{t}(\mathbf{s}))= 0,
\end{equation}
where $\delta \mathbf{v}_{t} (\mathbf{s})$ is the residual displacement vector field at location $\mathbf{s}$, and $\mathbf{v}_{t}(\mathbf{s})$ is the absolute vector field. 
At each iteration $\mathbf{v}_{t}(\mathbf{s})$ is updated and $I_t$ is warped according to the updated $\mathbf{v}_{t}$ field so that estimated residual displacements $\delta \mathbf{v}_{t}$ are expected to become progressively smaller.
These equations are coupled by the underlying physical geometry. 
For example, in the case of multi-view disparity estimation, the extrinsic camera parameters provide a coupling between $\delta \mathbf{v}_{t}(\mathbf{s})$ and the geometry of the scene. 
If multiple scales are being considered, different scales of the same image-pair will have the same residual displacement vector fields. 
Another example in optical flow is the loose coupling of consecutive frame pairs based on an acceleration term~\cite{JLGray_OpticalFlow2022}. 

In a variational framework, it is convenient to apply a local linearisation to~\eqref{eqn:Generalised_Brightness_Constancy}. 
We write the linearised equation as, 
\begin{equation}\label{eqn:Genearlised_Local_Linear}
    \langle \nabla I_{r, t} (\mathbf{s}), \delta \mathbf{v}_{t}(\mathbf{s}) \rangle + \delta I_{r, t} (\mathbf{s}) = 0,
\end{equation}
where $\nabla I_{r, t} (\mathbf{s})$ is the spatial gradient at location $\mathbf{s}$ averaged between images $r$ and $t$. 
Specifically,
\begin{equation}
    \nabla I_{r, t} (\mathbf{s}) = \frac{\nabla I_r (\mathbf{s}) + \nabla I_t (\mathbf{s})}{2},
\end{equation}
where $\nabla I_i (\mathbf{s})$ is the spatial gradient vector of the $i$th image at location $\mathbf{s}$. 
The $\delta I_{r, t} (\mathbf{s})$ term refers to the difference between images $r$ and $t$, i.e., $I_{t}(\mathbf{s}) - I_{r}(\mathbf{s})$. 

We write the energy minimisation problem as the sum of a brightness constancy constraint (the data term) and a smoothness constraint (the regularisation term). Specifically,
\begin{equation}\label{eqn:Gen_optimization}
E = E_{d} + \alpha E_{r},
\end{equation}
where $E_d$ is the data term, $E_r$ is the regularisation term and $\alpha$ is the regularisation parameter. 
This work primarily focuses on the data term, $E_d$, which is based on the linearised brightness constancy constraint,
\begin{equation}
    E_d = \iint_\Omega \sum_{t} \left( \langle \nabla I_{r, t} (\mathbf{s}), \delta \mathbf{v}_{t}(\mathbf{s}) \rangle + \delta I_{r, t} (\mathbf{s}) \right)^2 d\mathbf{s}.
\end{equation}
This quadratic formulation can be readily replaced by one involving a robust cost function such as the commonly used $L^1$ norm. 
However, this paper focuses on addressing issues with the local linearisation~\eqref{eqn:Genearlised_Local_Linear}. 
Therefore, we focus on the $L^2$ case, noting that the formulations developed can be readily generalised to robust norms, by using methods such as iteratively reweighted least squares (IRLS)~\cite{Welsch_IRLS_1977}.

\begin{figure}[!t]
\centering
    \begin{tikzpicture}[thick]
    \shadedraw[] (-3,-3) rectangle (3, 3);
    \filldraw[fill=yellow, opacity=0.25] (0, 0) circle (2);
    \draw (0, 0) circle (2);
    \coordinate[label = below right: $\nabla I_{0}(\mathbf{s})$] (O) at (0,0);
    \coordinate[label = below right: $\nabla I_{1}(\mathbf{s})$] (A) at (75:0.7);
    \coordinate[label = below right: $\nabla I_{2}(\mathbf{s})$] (B) at (75:1.4);
    \coordinate[label = below right: $\nabla I_{3}(\mathbf{s})$] (C) at (245:1.5);
    \coordinate[label = below right: $\nabla I_{4}(\mathbf{s})$] (D) at (150:2.0);
    \coordinate[label = below:$D(\mathbf{s})$] (Z) at (270:2);
    \coordinate[label = below: $\Omega$] (Y) at (270:3);
    \coordinate[] (X) at (180:2);
    \coordinate[label = below:$\delta \mathbf{v}_{\max}(\mathbf{s})$] (W) at (180:1);
    \node at (O)[circle,fill,inner sep=1pt]{};
    \node at (A)[circle,fill,inner sep=1pt]{};
    \node at (B)[circle,fill,inner sep=1pt]{};
    \node at (C)[circle,fill,inner sep=1pt]{};
    \node at (D)[circle,fill,inner sep=1pt]{};
    \draw[Stealth-Stealth] (O) -- (X);
    \draw[Stealth-Stealth, dotted] (O) -- (B);
    \end{tikzpicture}
\caption{A diagram of the region $D(\mathbf{s})$ in the image domain $\Omega$. One can think of the individual gradients $\nabla I_{i}(\mathbf{s})$ as observations of the underlying gradient of the image surface $\nabla I(\mathbf{s})$. 
For the local linearisation to be valid between $I_0$ and $I_2$, the gradient must be constant along the dotted line. 
To assess the validity of the linearisation, we can look at the gradients $\nabla I_{0}(\mathbf{s})$, $\nabla I_{1}(\mathbf{s})$ and $\nabla I_{2}(\mathbf{s})$; if any of these gradients are different, then the linearisation is invalid.}
\label{fig:Dq_Diagram}
\end{figure}
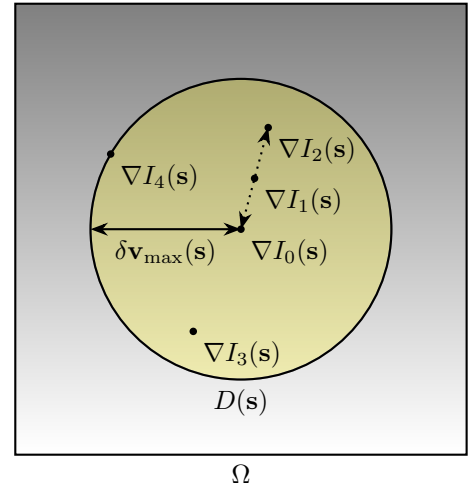

The local linearisation of the brightness constancy constraint~\eqref{eqn:Genearlised_Local_Linear} is invalid when $\delta \mathbf{v}_{t}(\mathbf{s})$ is large and when the image data contains high-frequency content. 
This is because~\eqref{eqn:Genearlised_Local_Linear} assumes that the gradient on the image surface $\nabla I(\mathbf{s})$ is constant over a region $D(\mathbf{s})$ around $\mathbf{s}$, which is large enough to contain $\delta \mathbf{v}_{r,t}(\mathbf{s})$ as shown in Figure~\ref{fig:Dq_Diagram}. 

We propose using Gradient Consistency information to assess the validity of~\eqref{eqn:Genearlised_Local_Linear}, weighting the data term accordingly. 
That is,
\begin{multline} \label{eq:weighted_l2_data_term}
        E_d = 
        \\ \iint_\Omega \sum_{t} W_{r, t}(\mathbf{s})( \langle \nabla I_{r, t} (\mathbf{s}), \delta \mathbf{v}_{t}(\mathbf{s}) \rangle + \delta I_{r, t} (\mathbf{s}))^2 d\mathbf{s},
\end{multline}
where $W_{r, t}(\mathbf{s})$ are the weights for each image pair. 
Note that weighting the data term in this way complements the use of a robust cost function. 
A robust cost function essentially down-weights violations of the brightness constancy constraint~\eqref{eqn:Generalised_Brightness_Constancy};
whereas, we use Gradient Consistency information to downweight areas where the linearisation~\eqref{eqn:Genearlised_Local_Linear} is not valid.

Both $I_r(\mathbf{s})$ and $I_t(\mathbf{s})$ exist on the image surface $D(\mathbf{s})$ separated by $\delta \mathbf{v}_{t}(\mathbf{s})$, as shown in Figure \ref{fig:Dq_Diagram}.
For~\eqref{eqn:Genearlised_Local_Linear} to hold, the gradient of the image surface along the vector $\delta \mathbf{v}_{t}(\mathbf{s})$ between $I_r(\mathbf{s})$ and $I_t(\mathbf{s})$ must be constant. 
The key observation that underpins Gradient Consistency is that~\eqref{eqn:Genearlised_Local_Linear} cannot hold unless $\nabla I_r(\mathbf{s})$ and $\nabla I_t(\mathbf{s})$ are equal. 
This property applies to each image pair, offering substantial insight into the relative validity of~\eqref{eqn:Genearlised_Local_Linear} for each image pair. 
In fact, in the multi-view case where co-linear views are available, there is additional evidence as to whether the gradient along the vector $\delta \mathbf{v}_{r,t}(\mathbf{s})$ is constant or not.
This situation is depicted in Figure~\ref{fig:Dq_Diagram}, where $I_1$ lies in between $I_0$ and $I_2$;
to the extent any of the gradients are different, the linearisation between $I_0(\mathbf{s})$ and $I_2(\mathbf{s})$ is not valid.

In Section~\ref{sec:GCM}, we formalise these rough ideas into an analytical method that we call the Gradient Consistency Model.


\section{Scale Consistency}\label{sec:ScaleConsistency}

In addition to multiple views, multiple scales also provide a useful source of evidence for the displacement field; these scales are often necessary when estimating large displacements.
The brightness constancy constraint can be formulated at each scale and we use Scale Consistency to determine their relative validity. 

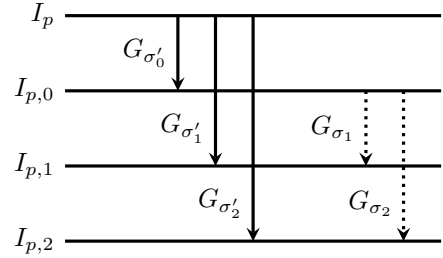
\begin{figure}[!t]
    \centering
    \begin{tikzpicture}
      \def\spacing{1}
      \def\hspace{0.5}
    
      \foreach \i/\label in {0/$I_p$, 1/$I_{p, 0}$, 2/$I_{p, 1}$, 3/$I_{p,2}$} {
        \draw[very thick] (0, -\i*\spacing) -- (5, -\i*\spacing)
        node[anchor=east] at (0, -\i*\spacing) {\label};
      }
    
      \foreach \i/\label in {1/$G_{\sigma_0^\prime}$, 2/$G_{\sigma_1^\prime}$,3/$G_{\sigma_2^\prime}$}{
        \draw[->,>=stealth, very thick] (\i * \hspace + 1,0) -- (\i * \hspace + 1, -\i*\spacing)
        node[anchor=east] at (\i * \hspace + 1, -\i*\spacing+\hspace) {\label}
        ;
      }
    
      \foreach \i/\label in {2/$G_{\sigma_{ 1}}$,3/$G_{\sigma_{2}}$}{
        \draw[->,>=stealth, dotted, very thick] (\i * \hspace + 3,-1) -- (\i * \hspace + 3, -\i*\spacing)
        node[anchor=east] at (\i * \hspace + 3, -\i*\spacing+\hspace) {\label}
        ;
      }
    \end{tikzpicture}
    \caption{This diagram illustrates how we consider the scales in this treatment. $I_p$ is the original image data from the $p$\textsuperscript{th} image. We filter the original image data using 
    $G_{\sigma_0^\prime}$ to produce the reference scale, $q=0$ of the $p$\textsuperscript{th} image, $I_{p,0}$.
    Different scales, $q \neq 0$ can be produced by filtering $I_p$ with $G_{\sigma_q^\prime}$ or by filtering $I_{p,0}$ with $G_{\sigma_q}$.
    }
    \label{fig:scale_diagram}
\end{figure}

Since we have introduced multiple scales, we augment our subscript notation for the images; 
writing $I_{t,q}$ and $I_{r,q}$ to represent target view $t$ and reference view $r$ at scale $q$. 
In this work, each $I_{p,q}$ is derived by convolving the original image, $I_p$ with a Gaussian filter, $G_{\sigma_q^\prime}$, as shown in Figure~\ref{fig:scale_diagram}.
Accordingly, gradients and image differences at scale $q$ are all computed with respect to the scale parameter $\sigma_{q}^\prime$, as
\begin{equation}
    \delta I_{t, q}(\mathbf{s}) = (G_{\sigma_q^\prime} * (I_{t} - I_{r}))(\mathbf{s}),
\end{equation}
and
\begin{equation} \label{eqn:nabla_I_defn}
    \nabla I_{t,q}(\mathbf{s}) =
    \frac{1}{2}
    \begin{bmatrix} 
    (G^\prime_{s_1, \sigma_q^\prime} * (I_{t} + I_{r}))(\mathbf{s})\\
    (G^\prime_{s_2, \sigma_q^\prime} * (I_{t} + I_{r}))(\mathbf{s})
    \end{bmatrix},
\end{equation}
where, $G^\prime_{s_i, \sigma_q^\prime}$ is the Derivative of Gaussian operator in the $s_{i}$ direction.

We could set $\sigma^\prime_0$ to zero, meaning that the top scale would consist of the original images. 
However, it is convenient to incorporate low-pass filtering for all scales, as shown in Figure~\ref{fig:scale_diagram}. 
Therefore, we set,
\begin{equation}\label{eqn:sigma_q_defn}
    \sigma_q = {2^q}/{\sqrt{2}}.
\end{equation}

In a multi-scale approach, the data term consists of the finest scale $q=0$ and several coarser scales, where $q \geq 1$.
The linearised brightness constancy constraint for one such coarse scale can be written as,
\begin{equation}\label{eqn:Filt_BCC_Data}
         J_B(\mathbf{s}) =
         \left\langle (G_{\sigma_q} * \nabla I_{t, 0} ) (\mathbf{s}), \delta \mathbf{v}_{t} (\mathbf{s}) \right\rangle + (G_{\sigma_q} * \delta I_{t, 0})(\mathbf{s}).
\end{equation}
However, displacement fields that are consistent with~\eqref{eqn:Filt_BCC_Data} at scale $q$ are not necessarily consistent with other scales; 
this is what causes scale inconsistency.
In contrast, filtered versions of~\eqref{eqn:Genearlised_Local_Linear} as presented in~\eqref{eqn:Filt_BCC} would be locally consistent with the brightness constancy constraint.
\begin{equation}
\label{eqn:Filt_BCC}
    J_A(\mathbf{s}) = 
    (G_{\sigma_q}* \langle \nabla I_{t, 0}, \delta \mathbf{v}_{t} \rangle)(\mathbf{s}) + (G_{\sigma_q} * \delta I_{t, 0}) (\mathbf{s}).
\end{equation}
We account for scale inconsistency by incorporating a bound on the local squared difference between $J_A(\mathbf{s})$ and $J_B(\mathbf{s})$ into our overall Gradient Consistency model, as formalized in Section~\ref{sec:GCM}.


\section{Physical Coupling of Displacement Fields} \label{sec:Physical_Coupling}

In multi-view disparity estimation, the $\delta \mathbf{v}_{t}(\mathbf{s})$ terms are coupled by the epipolar constraints. 
The simplest case is a planar camera array where all cameras have identical orientation and each target view has a baseline $\mathbf{B}_t$ between itself and the reference view. 
We focus our attention on this case, because in this paper, the experiments are performed with the camera extrinsic parameters set up in this way.
In this case, the coupling between the $\delta \mathbf{v}_{t}(\mathbf{s})$ terms is quite simple; each $\delta \mathbf{v}_{t}(\mathbf{s})$ is a based on the residual reciprocal depth (sometimes called the residual normalised disparity) $\delta w(\mathbf{s})$ and the baseline $\mathbf{B}_{t}$. Specifically, 
\begin{equation} \label{eqn:Displacement_Disparity_relation}
    \delta \mathbf{v}_{t}(\mathbf{s}) = \mathbf{B}_t \cdot \delta w(\mathbf{s}).
\end{equation}

Accordingly, the brightness constancy constraint can be written as,
\begin{equation}\label{eqn:brightConst}
     I_{r,q}(\mathbf{s}) - I_{t,q}(\mathbf{s} + \mathbf{B}_{t} \cdot \delta {w}(\mathbf{s})) = 0.
\end{equation}
Therefore, the brightness constancy constraint can be linearised as,
\begin{equation}\label{eqn:brightConstLin}
     g_{t,q}(\mathbf{s}) \delta w(\mathbf{s}) + \delta I_{t,q}(\mathbf{s}) = 0,
\end{equation}
where, 
\begin{equation} \label{eqn:grad_defn}
    g_{t,q}(\mathbf{s}) = \langle \nabla I_{t,q}(\mathbf{s}) , \mathbf{B}_{t} \rangle.
\end{equation}

\section{Gradient Consistency Model}\label{sec:GCM}
The Gradient Consistency Model is an analytical model for determining the weights in~\eqref{eq:weighted_l2_data_term} by considering Gradient Consistency and Scale Consistency information.

We derive the weights for the GCM based on the Euler-Lagrange equations that minimise~\eqref{eq:weighted_l2_data_term}, in the context of multi-view disparity estimation with the coupling discussed in Section~\ref{sec:Physical_Coupling}. 
To simplify this derivation, we ignore the regularisation term by essentially setting the regularisation parameter $\alpha$ to $0$.
Accordingly, the Euler-Lagrange equations can be written as,
\begin{equation}
\label{eqn:ral_EL_eqn}
\mathbf{0}  =
    \sum_{t,q} W_{t, q}(\mathbf{s}) g_{t,q}(\mathbf{s}) (g_{t,q}(\mathbf{s}) \delta w(\mathbf{s}) + \delta I_{t,q}(\mathbf{s})).
\end{equation}
The solution to~\eqref{eqn:General_EL_eqn} is essentially a weighted sum of the $\delta I_{t, q}(\mathbf{s})$ terms,
\begin{equation} \label{eqn:Data_Term_estimate}
    \delta {w}^\prime(\mathbf{s}) = -\frac{ \sum_{t,q} W_{t, q}(\mathbf{s}) g_{t,q}(\mathbf{s}) \delta  I_{t,q}(\mathbf{s})} 
{\sum_{t,q} W_{t, q}(\mathbf{s})g^2_{t,q}(\mathbf{s})}.
\end{equation}

Suppose that an imaging noise term, $\mathcal{N}_{t,q}(\mathbf{s})$ is applied to each $\delta  I_{t,q}(\mathbf{s})$; this will perturb the solution of~\eqref{eqn:General_EL_eqn}. 
We can write the perturbed solution as,
\begin{equation}\label{eqn:Noisy_solution}
    \delta {w}^*(\mathbf{s}) = \delta {w}^\prime(\mathbf{s}) - \frac{ \sum_{t, q} W_{t, q}(\mathbf{s}) g_{t,q}(\mathbf{s}) \mathcal{N}_{t, q}(\mathbf{s}) } 
{\sum_{t,q} W_{t, q}(\mathbf{s}) g^2_{t,q}(\mathbf{s})}.
\end{equation}
As shown in Appendix~\ref{appendix:weights}, we can minimise the expected squared error in the solution by selecting weights such that, 
\begin{equation} \label{eqn:Weights_Proportional}
   W_{t,q}(\mathbf{s}) = \frac{Z}{E[\mathcal{N}^2_{t,q}(\mathbf{s})]},
\end{equation}
where $Z$ is an arbitrary positive constant.

The result in \eqref{eqn:Weights_Proportional} considers only the effect of uncorrelated imaging noise on the choice of weights, whereas we are primarily interested in the effect of gradient and scale inconsistency. Directly modeling the effect of noise in the $g_{t,q}(\mathbf{s})$ terms of equation \eqref{eqn:Noisy_solution} with a view to minimizing the expected squared error in the solution turns out to be problematic. Therefore, we resort to a simpler strategy which only considers the effect of gradient noise on the numerator of equation \eqref{eqn:Noisy_solution}. Specifically, we approximate the effect of gradient inconsistency as an equivalent imaging noise and include it in $\mathcal{N}^2_{t, q}(\mathbf{s})$. 
We also include a scale inconsistency term in $\mathcal{N}^2_{t, q}(\mathbf{s})$, as well as an image acquisition noise term. Specifically,
\begin{equation}\label{eqn:exp_noise_pwr}
    \mathcal{N}^2_{t, q}(\mathbf{s}) = \mathcal{G}_{t, q}^2(\mathbf{s}) \delta {w}^2_{q, e}(\mathbf{s}) + \mathcal{O}_{t,q}^2(\mathbf{s}) + \frac{\epsilon^2}{4 \pi \sigma_{q}^2}.
\end{equation}
The $\mathcal{G}_{t, q}^2(\mathbf{s})$ term represents gradient inconsistency. As per Section~\ref{sec:Intro_to_GC}, it is primarily based on the difference between the gradients of $I_t$ and $I_r$ at scale $q$ and location $\mathbf{s}$, 
\begin{equation}
\label{eqn:Grad_diff}
    \mathcal{G}_{t,q}(\mathbf{s}) = \frac{1}{2} \left \langle  
    \begin{bmatrix} 
    (G^\prime_{s_1, \sigma_q} * (I_{t} - I_{r}))(\mathbf{s})\\
    (G^\prime_{s_2, \sigma_q} * (I_{t} - I_{r}))(\mathbf{s})
    \end{bmatrix} , \mathbf{B}_{t} \right \rangle.
\end{equation}
We take the inner product of the baseline $\mathbf{B}_{t}$ and the difference between the two gradients, because $\nabla I_{t,q}(\mathbf{s})$ itself projected onto $\mathbf{B}_{t}$ in~\eqref{eqn:grad_defn}. Therefore, any inconsistencies will also be projected onto $\mathbf{B}_{t}$ using an inner product.

The $\delta {w}^2_{q,e}(\mathbf{s})$ term is an upper bound for the estimated error in the current absolute disparity, ${w}(\mathbf{s})$. 
\begin{equation}\label{eqn:Delta_w_defn}
    \delta w_{q, e}^2(\mathbf{s}) = \frac{\frac{\epsilon^2}{4 \pi \sigma_q^2} + \sum_{t,q} \delta I^2_{t,q}(\mathbf{s})}{\sum_{t, q} g^2_{t,q}(\mathbf{s}) + \epsilon} + \text{var}_{\sigma_c}(w_{\text{prev}}(\mathbf{s})),
\end{equation}
where $\delta w_{\text{prev}}(\mathbf{s})$ is $\delta w(\mathbf{s})$ in the previous iteration, and the $\text{var}_{\sigma_c}()$ operator is defined as,
\begin{equation} \label{eqn:w_std}
    \text{var}_{\sigma_c}(x(\mathbf{s})) = {(G_{\sigma_c} * x^2)(\mathbf{s}) - (G_{\sigma_c} * x)^2(\mathbf{s})}.
\end{equation}
The first term in~\eqref{eqn:Delta_w_defn} provides an upper bound on the error of $w_\text{prev}(\mathbf{s})$ based on the image data.
Whereas, the second term is the local variance in $w_\text{prev}(\mathbf{s})$ which also provides an upper bound based on smoothness assumptions. 

In~\eqref{eqn:exp_noise_pwr}, the $\mathcal{O}_{t, q}^2(\mathbf{s})$ term is a scale inconsistency term based on the difference between $J_A(\mathbf{s})$ and $J_B(\mathbf{s})$ as discussed in Section~\ref{sec:ScaleConsistency}.
In Appendix~\ref{appendix:Spatial} we show that an upper bound for $\mathcal{O}_{t, q}^2(\mathbf{s})$ in a multi-view disparity estimation context is,
\begin{multline}\label{eqn:SpatialInconsistency_U}
         \mathcal{O}^2_{t,q}(\mathbf{s}) \approx \left(
        \iint_\Omega G_{\sigma_q}(\boldsymbol{\tau}) g_{t, 0}^2(\mathbf{s} - \boldsymbol{\tau}) d \boldsymbol{\tau} \right)\\
        \cdot \left(
        \iint_\Omega G_{\sigma_q}(\boldsymbol{\tau}) \widetilde{\delta w}^2(\mathbf{s} - \boldsymbol{\tau}) d \boldsymbol{\tau} \right),
\end{multline}
where $\widetilde{\delta w}(\mathbf{s})$ is the estimated residual disparity at scale $q=0$.
Since $\widetilde{\delta w}(\mathbf{s})$ is unknown at the time of calculation, we approximate it in a point-wise fashion based on the image data at scale $q=0$,
\begin{equation} \label{eqn:tilde_dw}
    \widetilde{\delta w}(\mathbf{s}) = \frac{\sum_t |\delta I_{t, 0} (\mathbf{s})|}{\sum_t |g_{t, 0}(\mathbf{s})| + \epsilon}.
\end{equation}
This approximation takes into account each image pair and each image difference at scale $q=0$, which is necessary since the $\widetilde{\delta w}(\mathbf{s})$ is common to all views and scales. 


The $\frac{\epsilon^2}{4 \pi \sigma_{q}^2}$ term in~\eqref{eqn:exp_noise_pwr} represents image acquisition noise power and its dependence on the scale $q$;
it also prevents $N_{t,q}^2(\mathbf{s}) \rightarrow 0$, which is necessary for the stability of the model.
The derivation for this term is included in Appendix~\ref{appendix:ImagingNoise}. We arbitrarily set $\epsilon = 2 \times 10^{-4}$. As shown in Table~\ref{table:epsilon_changes}, it turns out that the GCM is substantially insensitive to changes in $\epsilon$.

As discussed in Section~\ref{sec:Intro_to_GC}, in the multi-view case there is additional evidence available about the consistency of gradients along $\delta \mathbf{v}_{t}(\mathbf{s})$ between $I_t(\mathbf{s})$ and $I_r(\mathbf{s})$.
In fact, views in between $I_t(\mathbf{s})$ and $I_r(\mathbf{s})$ must also have consistent gradients for~\eqref{eqn:Genearlised_Local_Linear} to hold over the length of $\delta \mathbf{v}_{t}(\mathbf{s})$.
To this end, we impose a monotonicity constraint on the Gradient Consistency Model weights. The baseline vector space is divided into 8 sectors, $\{S_k\}^7_{k = 0}$.
We assign a view $\mathcal{V}_t$, whose normalized location is $\mathbf{B}_t$, to sector $S_k$ if the angle $\angle \mathbf{B}_t$ is within that sector's angle range, i.e., $k \cdot \pi/4 \leq \angle\mathbf{B}_t \leq (k + 1) \cdot \pi/4$.  The weight $W_{t, q}(\mathbf{s})$ for view $t$ at scale $q$ has a monotonicity constraint applied to it. Specifically, 
\begin{equation}
    W_{t, q}(\mathbf{s}) = 
    \min_{\substack{n \\ \vert\mathbf{B}_n\vert \leq \vert\mathbf{B}_t\vert \text{ \& } \mathcal{V}_n, \mathcal{V}_t \in S_k}}
    \left( 
        \frac{Z}{E[\mathcal{N}_{n, q}^2(\mathbf{s})]}
    \right).
\end{equation}


The GCM discussed thus far has been formulated for a multi-view disparity estimation context with co-planar cameras that have identical orientation. 
However, the Gradient Consistency Model can actually be generalised to other contexts, so long as there are linear couplings between the displacement fields and an underlying scalar field, i.e.
\begin{equation}
    \delta \mathbf{v}_{t, q}(\mathbf{s}) = \mathbf{c}_{t, q}(\mathbf{s}) \cdot \delta w(\mathbf{s}) + \mathbf{d}_{t,q}(\mathbf{s}).
\end{equation}
Of course, the $\mathcal{G}_{t, q}(\mathbf{s})$, $\mathcal{O}_{t, q}(\mathbf{s})$ and $\delta w_{q,e}^2(\mathbf{s})$ terms
must be adapted to the specific situation where the GCM is applied.

\section{Variational Disparity Estimation} \label{sec:VariationalModel}

Our discussion of the Gradient Consistency Model in Section~\ref{sec:GCM} has ignored  regularisation. 
In practice, a regularisation term is necessary to realise a variational framework.
Therefore, we use~\eqref{eqn:Gen_optimization} as the total cost function.
For the regularisation term itself, we select the well-known Total Variation regulariser.
This can be written as,
\begin{equation} \label{eqn:RegTerm}
    E_{r} = \iint_\Omega  \psi( \nabla w(\mathbf{s})) d\mathbf{s},
\end{equation}
where $\psi()$ is the $L^1$ norm and $w(\mathbf{s})$ is the disparity.
The data term, $E_d$ is written as,
\begin{multline}
\label{eqn:Robust_Data_Term}
    E_{d} = \\ \iint_\Omega \sum_{t, q} W_{t,q}(\mathbf{s}) 
    \psi ( I_{r,q}(\mathbf{s}) - I_{t,q}(\mathbf{s} + \mathbf{B}_{t} \cdot \delta{w}(\mathbf{s})) ) d\mathbf{s},
\end{multline}
where the $W_{t,q}(\mathbf{s})$ weights are determined using the GCM.
We employ the standard gradient-based linearisation along with an IRLS scheme to transform~\eqref{eqn:Robust_Data_Term} into,
\begin{multline} \label{eqn:Linear_Data_term}
    E_{d} = \\
    \iint_\Omega \sum_{t, q} W_{t,q}(\mathbf{s}) R_{t,q}(\mathbf{s})
    (g_{t,q}(\mathbf{s}) \delta {w}(\mathbf{s}) + \delta I_{t,q}(\mathbf{s}))^2 d\mathbf{s},
\end{multline}
where $R_{t,q}(\mathbf{s})$ are the weights applied by the IRLS scheme. 

We use the $L^1$ norm for both the regulariser and data term, because it is robust and well understood.

\section{Single Scale Results}\label{sec:Single_Scale_Results}
\subsection{4D Lightfield Dataset}

We evaluate the performance of the Gradient Consistency Model on a synthetic 4D Lightfield Dataset~\cite{Honauer2016} by comparing the GCM with two other methods: the progressive inclusion of views approach~\cite{JLGray2021}, with only one scale used and 4 stages of including views; and the naive approach of assigning all locations in all views a weight $W_{t,q}(\mathbf{s}) = 1$.

We compare each method across 24 scenes, from the stratified, additional and training sets from~\cite{Honauer2016}. For each scene, we use 17 of the 81 views arranged in a cross-hair shape. 

Each method uses the $L^1$ loss function as per Section~\ref{sec:VariationalModel}, realised using an iteratively reweighted least squares scheme~\cite{Welsch_IRLS_1977}. 
Each least squares solution is determined using the conjugate gradient technique~\cite{Hestenes1952}. 
Aside from this, 
the progressive inclusion of views and the naive approach do not apply any weighting schemes to the data; each target view is considered to be equal. 
In contrast, the GCM directly applies a weighting scheme to the image data.

For the purposes of this test, the progressive inclusion of views approach begins with 5 views in the same cross-hair shape. 
To ensure stability and progress through the progressive inclusion of views in a timely fashion, we apply a limiting policy to the change in disparity $\delta w(\mathbf{s})$ from one IRLS stage to another. 
If $|\delta w(\mathbf{s})|$ is larger than the limit $M$, we clip $\delta w(\mathbf{s})$, such that $|\delta w(\mathbf{s})| \leq M$. 
If no limiting occurs at a given stage, we move to the next stage and include 4 more views.
We then repeat the process until the final stage, when we have included all the views. At this point, we allow the algorithm to run until it converges or reaches the maximum number of IRLS stages. We define $M$ as, 
\begin{equation}\label{eqn:Limit_Scheme}
    M = {1}/{|\mathbf{B}_{max}|},
\end{equation}
where $\mathbf{B}_{max}$ is the baseline of the furthest view from the reference view.

For convenience, the baselines $\mathbf{B}_{t}$ and disparity fields $\delta w(\mathbf{s})$ are normalized so that the nearest pair of views have $|\mathbf{B}_{t}| = 1$.

After each stage, we use a simple 5x5 median filter to remove outliers from the disparity field.

The strategies of median filtering and limiting the displacement field are commonly employed strategies in variational frameworks~\cite{Sun_2010}. 

All methods perform warping inside the IRLS scheme, prior to each reweighting stage. This is done to allow for more straightforward comparisons between the methods.

Inevitably, some disparity vectors will point outside the image boundaries in target views, presenting a problem during warping; we would be attempting to warp data that doesn't exist. 
We address this by copying data from the reference view into the warped target view to fill those areas which require data from outside the image.

Since the methods are distinct, the best choice of the regularisation parameter $\alpha$ may not be the same for all methods. 
For this reason, we adjust $\alpha$ within the 6-fold validation framework on the training, stratified and additional scenes in the 4D Lightfield Dataset~\cite{Honauer2016}. 

\begin{figure}[!t]
\centering
\includegraphics[width=8.5cm]{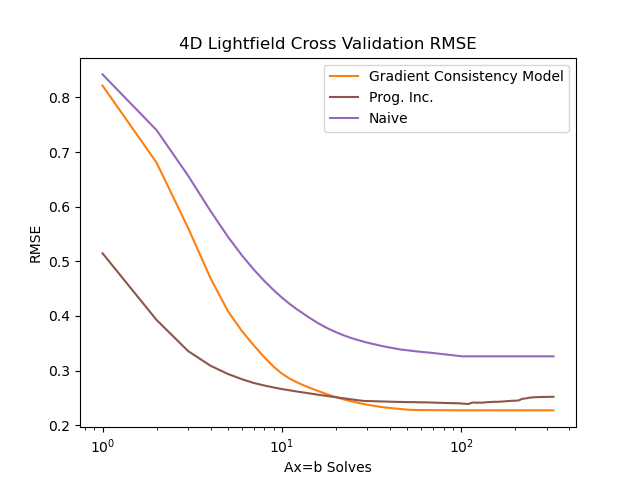}
\caption{Plots of the RMSE vs. the number of $A\mathbf{x} = \mathbf{b}$ solves (log scale) for the average of the 6-fold cross-validation on a synthetic 4D Lightfield Dataset \cite{Honauer2016}. Here we compare 
the naive approach (purple) with the Gradient Consistency Model (orange) and the progressive inclusion of views (brown).}
\label{fig:1_scale_plot}
\end{figure}

The results of the 6-fold cross-validation are shown in Figure \ref{fig:1_scale_plot} and Table \ref{table:5FoldValidation_1_scale}. In Figure \ref{fig:1_scale_plot}, we plot the RMSE on the $y$-axis against the number of $A\mathbf{x} = \mathbf{b}$ solves on the $x$-axis, which has a log scale to better view the early stages of each approach. 
We use the number of $A\mathbf{x} = \mathbf{b}$ solves on the $x$-axis because, ultimately all variational optical flow algorithms use local linearisation to convert the original objective into a set of equations to solve, after which the linearization is performed again. The elegance of converting a difficult problem into a set of linear equations must be weighed against the need to reformulate the equations, so it makes sense to compare the impact of different approaches to the problem with respect to the number of such reformulations that are required — this is what we mean by $A\mathbf{x} = \mathbf{b}$ “solves”.

\begin{table}[!t]
\caption{Final 6-fold Cross-Validation Results on the 4D Lightfield Dataset. 
} \label{table:5FoldValidation_1_scale}
\centering
\npdecimalsign{.}
\npthousandsep{}
\nprounddigits{5}
\begin{tabular}{c|n{1}{5}}
        \textbf{Method} & \textbf{RMSE}  \\ \hline
        
        Naive & 0.3260942820416670  \\ 

        Gradient Consistency Model & 0.22723282625 \\ 

        Progressive Inclusion of Views & 0.2519847182916670 \\ 
\end{tabular}
\npnoround
\end{table}

From the results shown in Figure \ref{fig:1_scale_plot} and Table \ref{table:5FoldValidation_1_scale}, a number of observations are clear. 
As expected, the naive approach performs worst, converging slowest and achieving a poor final RMSE. 
The progressive inclusion of views converges rapidly but fails to achieve the best final RMSE.
In fact, the progressive inclusion of views actually has reduced accuracy as further views are added. 
In contrast, the GCM achieves the best final result and after 30 solves, it has a better RMSE than the other approaches. 

The more rapid early convergence of the progressive inclusion of views is the result of the limiting policy applied to each stage. 
Since the progressive inclusion of views begins with only the innermost views, $|\mathbf{B}_{max}| = 1$ and therefore $M$ from~\eqref{eqn:Limit_Scheme} is $1$.
Accordingly, early solves are allowed by the limiting heuristic to take larger steps. 
However, in the case of the naive approach and the GCM, all views are initially considered, and so, $|\mathbf{B}_{max}| = 4$, resulting in $M = 0.25$.
Consequently, early solves take much smaller steps, when compared to the progressive inclusion of views.

\begin{figure}[!t]
\centering
\includegraphics[width=8cm]{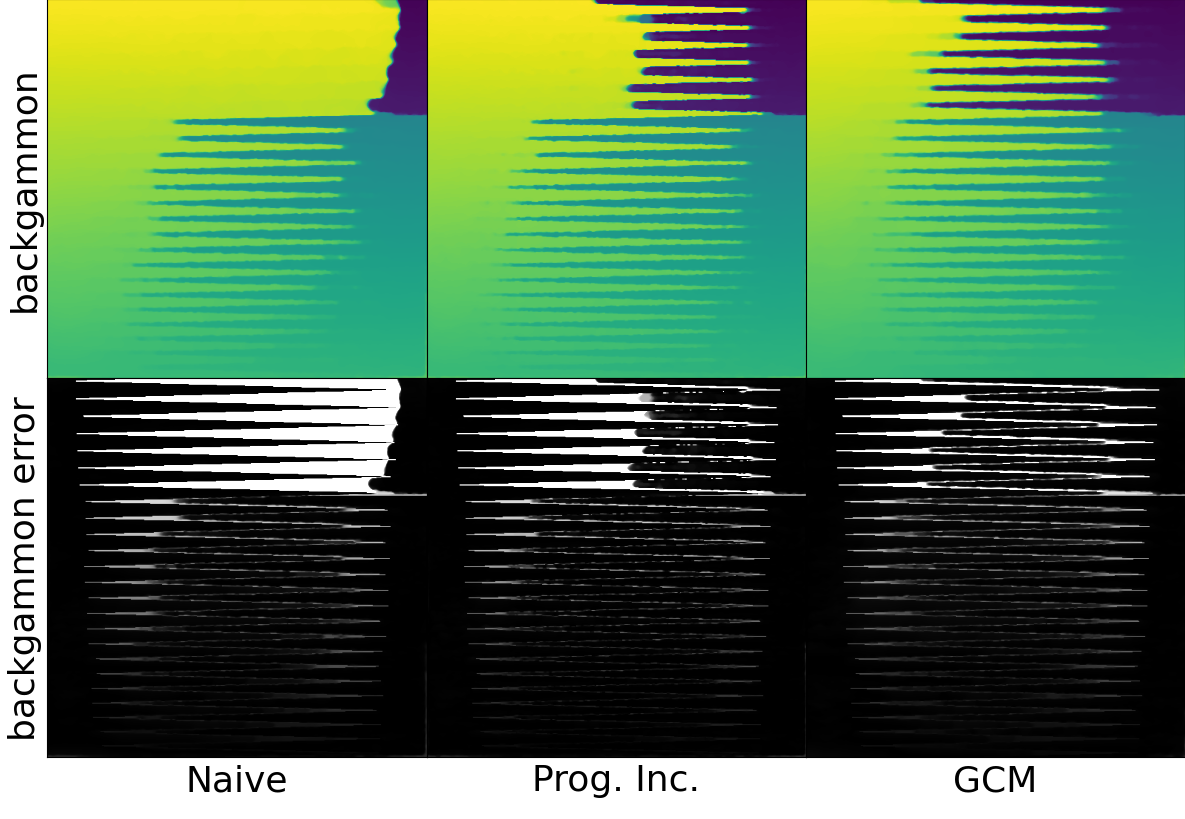}
\caption{On the top row, we have the estimated disparity fields from the three methods considered. All plots in the top row share a common colour scale. On the bottom row, we have the absolute error in the estimated disparity.  All plots in the bottom row share a common scale where, black is 0 and white is $(w_\text{max} - w_\text{min})/2$. Note that, $w_\text{max}$ is the maximum and $w_\text{min}$ is the minimum of the ground truth disparity field.}
\label{fig:backgammon_1scale}
\end{figure}

The better final result for the GCM seems to be largely due to better performance around discontinuities in the disparity field. 
The backgammon scene is an illustrative example of this; the disparity field has some very sharp discontinuities as shown in Figure \ref{fig:backgammon_1scale} and 
the error around the discontinuities is significantly lower for the GCM. 

The reason for the improved performance of the GCM around scene-object boundaries is that unlike the other approaches which treat each view equally, the GCM can downweight unreliable areas and up-weight more reliable areas. 
Near scene-object boundaries, the gradients will be less consistent, especially for views further apart. 
To mitigate this, the GCM favours the inner views in these areas which tend to better fit the linearised brightness constancy constraint.

\section{Multi-Scale Results} \label{sec:Eval}
\subsection{4D Lightfield Dataset} \label{ssec:4D_Lightfield}
In this section we expand the optimisation strategy to cover three spatial scales rather than one.

The primary comparison with these results is between a coarse-to-fine variational multi-view disparity estimation approach based on \cite{Tran2017Disparity} and the GCM. 
Both of these methods use the same robust cost functions in both the data term and the regularisation term. 
Both methods use three scales of data, since the maximum disparities have a magnitude less than 4. 
To understand the GCM better and to illustrate the influence of each term in the GCM, we modify it in two key ways and include them in the comparison. 
These are: setting $\mathcal{O}_{t,q}^2(\mathbf{s}) =0$ and setting $\mathcal{G}_{t,q}^2(\mathbf{s}) = 0$.

To ensure stability and progress through the coarse-to-fine scheme in a timely fashion, we apply a similar limiting scheme used in the progressive inclusion of views described in Section~\ref{sec:Single_Scale_Results}.
The only difference is that~\eqref{eqn:Limit_Scheme} is replaced with,
\begin{equation}\label{eqn:Limit_scheme_multi}
    M = {2^q}/{|\mathbf{B}_{max}|},
\end{equation}
where $q$ is the current scale.
Ultimately, the limiting and scheduling policy that one uses in a coarse-to-fine scheme is a choice. 
On one hand, if smaller values of $M$ are chosen, the algorithm will spend longer at coarse scales since it allows for smaller disparity changes at each $A\mathbf{x} = \mathbf{b}$ solve. On the other hand, if larger values of $M$ are selected, the algorithm will move to finer scales earlier. 
We use~\eqref{eqn:Limit_scheme_multi} because it is consistent with~\eqref{eqn:Limit_Scheme} in the $q=0$ case and moreover, we empirically observed that choosing a larger value of $M$ tends to reduce accuracy while choosing a smaller value of $M$ tends to unnecessarily slow convergence.

\begin{figure}[!t]
\centering
\includegraphics[width=8.5cm]{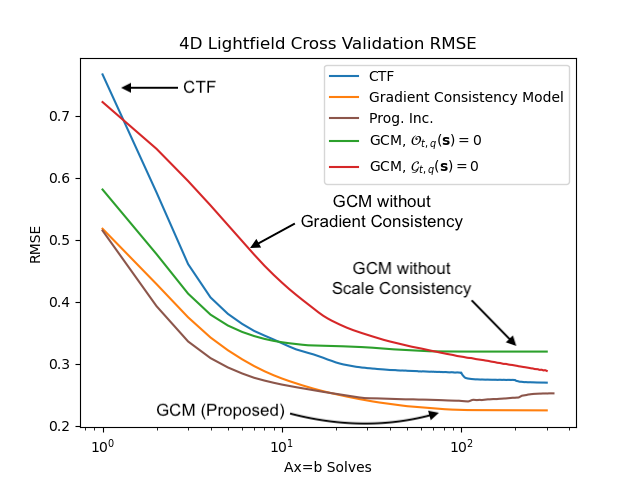}
\caption{Plots of the RMSE vs. the number of $A\mathbf{x} = \mathbf{b}$ solves (log scale) for the average of the 6-fold cross-validation on a synthetic 4D Lightfield Dataset \cite{Honauer2016}. Here we compare the coarse-to-fine approach (blue) based on \cite{Tran2017Disparity} with the Gradient Consistency Model (orange), the GCM with $\mathcal{O}_{t,q}^2(\mathbf{s}) = 0$, the GCM with $\mathcal{G}_{t,q}^2(\mathbf{s}) = 0$ (red) and the progressive inclusion of views (brown).}
\label{fig:Log_RMSE_plot}
\end{figure}

Similar to Section~\ref{sec:Single_Scale_Results}, we vary the regularisation parameter $\alpha$ using a 6-fold cross-validation framework. We plot the average RMSE value for each method after 6-fold cross-validation against the number of $A\mathbf{x} = \mathbf{b}$ solves in Figure \ref{fig:Log_RMSE_plot}. 
To better view the early stages of each approach, a log scale is used on the $x$-axis.
The final result of each approach is shown in Table \ref{table:5FoldValidation},

\begin{table}[!t]
\caption{Final 6-fold Cross-Validation Results on the 4D Lightfield Dataset.
} \label{table:5FoldValidation}
\centering
\npdecimalsign{.}
\npthousandsep{}
\nprounddigits{5}
\begin{tabular}{c|n{1}{5}}
        \textbf{Method} & \textbf{RMSE} \\ \hline
        Coarse to Fine & 0.26925784950000000 \\ 
    
        Gradient Consistency model & 0.224522 \\ 

        GCM, $\mathcal{O}_{t,q}(\mathbf{s})= 0$ & 0.3194290569583330 \\ 

        GCM, $\mathcal{G}_{t,q}(\mathbf{s})= 0$ & 0.288265 \\ 

\end{tabular}
\npnoround
\end{table}

There is a clear pattern with the results shown in Figure \ref{fig:Log_RMSE_plot};
the GCM achieves accurate results significantly earlier than the coarse-to-fine scheme.
In fact, the GCM only requires roughly 30-40 solves to achieve a similar RMSE to its final result, as opposed to 100-200 solves depending on the scene for the coarse-to-fine scheme.
Furthermore, at all points during the algorithm, the multi-scale GCM is more accurate than the coarse-to-fine scheme. 

\begin{figure}[!t]
\centering
\includegraphics[width=8.5cm]{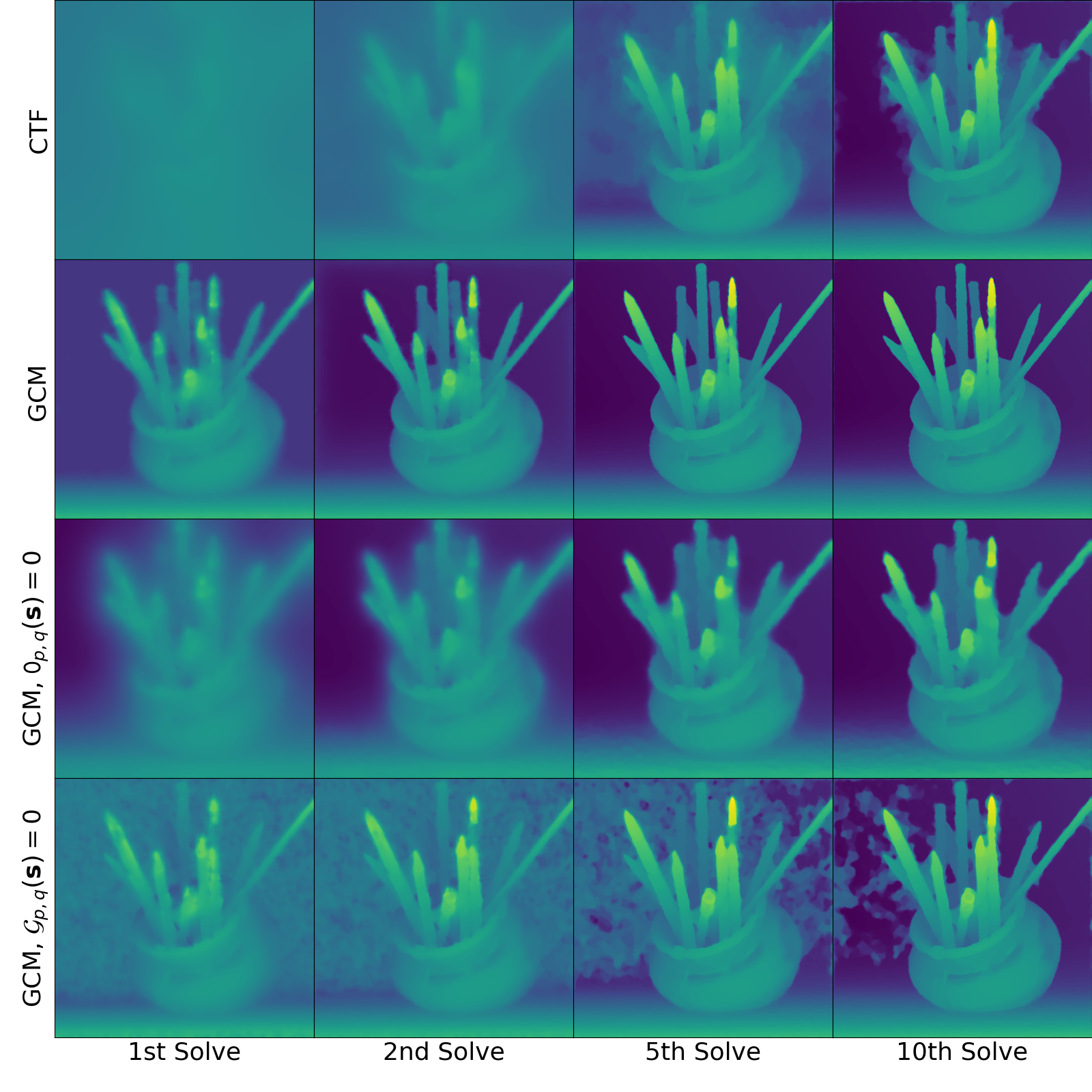}
\caption{In each column of this plot we have a disparity field plot generated after a given number of solves using several different approaches. All plots share a common colour scale.}
\label{fig:pens_progress}
\end{figure}

We can visualise the improved rate of convergence of the GCM in Figure \ref{fig:pens_progress}. By the 10th solve, the GCM has mostly converged. However, the coarse-to-fine scheme is still converging. This is because the coarse-to-fine scheme cannot use fine detail in the initial solves. 

\begin{figure}[!t]
\centering
\includegraphics[width=8.5cm]{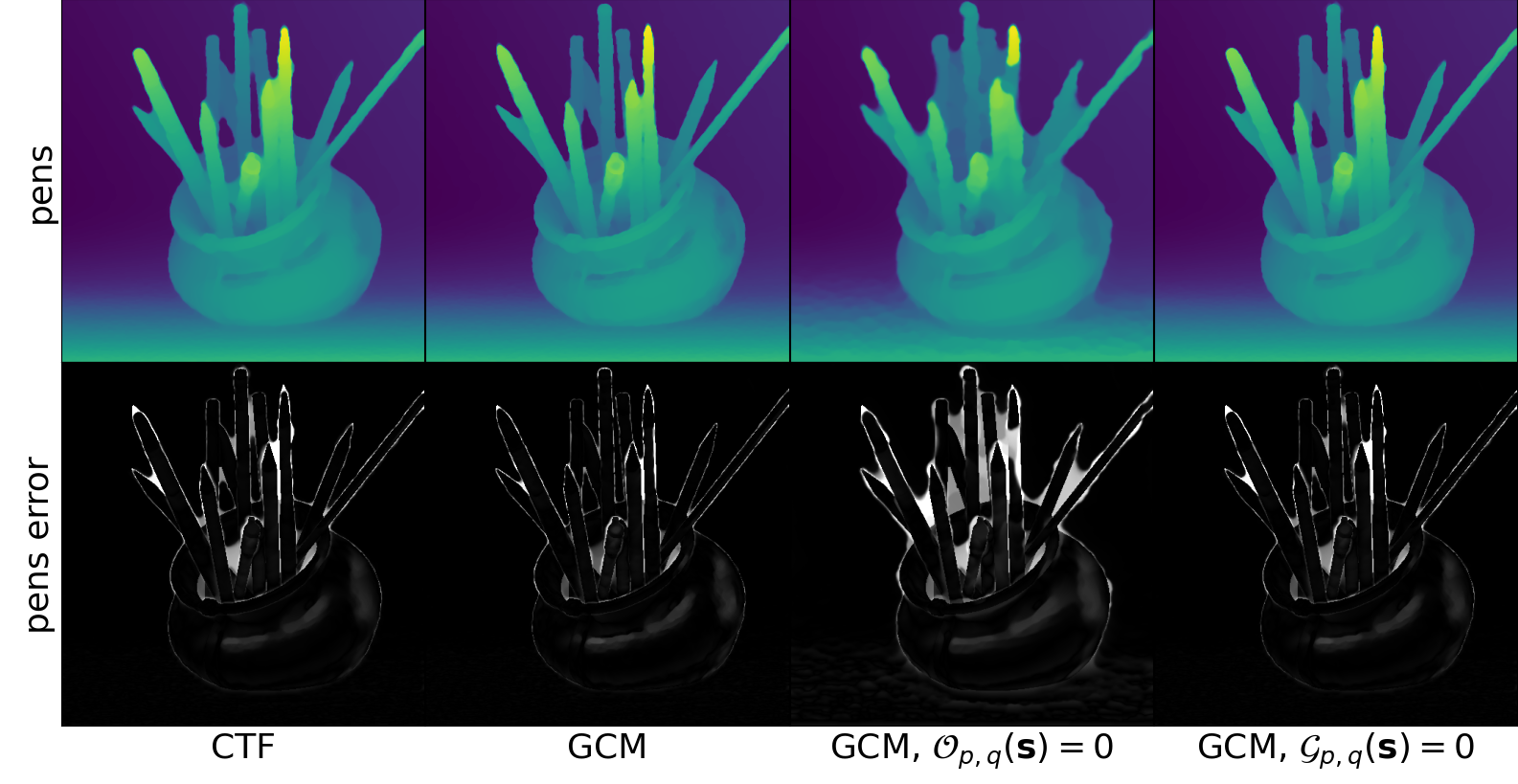}
\caption{On the top row, we have the final estimated disparity fields from the four methods considered. All plots in the top row share a common colour scale. On the bottom row, we have the absolute error in the estimated disparity. All plots in the bottom row share a common scale, where black is 0 and white is $(w_\text{max} - w_\text{min})/2$. Note that, $w_\text{max}$ is the maximum and $w_\text{min}$ is the minimum of the ground truth disparity field.
}
\label{fig:pens_3scale}
\end{figure}

Similar to the single scale results, much of the improvement in the GCM results seems to be due to improved performance around object boundaries. 
This is illustrated in Figure \ref{fig:pens_3scale}; 
the error bands near object boundaries are much smaller when using the multi-scale GCM instead of a coarse-to-fine framework. 
These results show that the GCM can be readily applied to three scales, producing improvements in accuracy and significant improvements in the rate of convergence. 

Both the $\mathcal{G}_{t,q}(\mathbf{s})$ and $\mathcal{O}_{t, q}(\mathbf{s})$ terms are crucial to the GCM. 
In fact, they perform key complementary roles; the $\mathcal{G}_{t, q}(\mathbf{s})$ term is highly important to the rate of convergence of the GCM and the $\mathcal{O}_{t, q}(\mathbf{s})$ term is important for the overall accuracy of the GCM.
This can be observed in Figure~\ref{fig:Log_RMSE_plot} which contains cases in which either either $\mathcal{G}_{t, q}(\mathbf{s})$ or $\mathcal{O}_{t, q}(\mathbf{s})$ is set to zero.

If we set $\mathcal{G}_{t, q}(\mathbf{s})$ to 0, convergence slows significantly.
In fact, even after 300 solves the disparity fields have not yet converged in all scenes, whereas all other approaches converge prior to that. 
An illustrative example of this is Figure~\ref{fig:pens_progress}, where we can observe that if $\mathcal{G}_{t, q}(\mathbf{s})$ is set to zero, the background of the scene converges much more slowly. 
This is because when $\mathcal{G}_{t, q}(\mathbf{s})$ is zero, the GCM excessively favours fine scale data and down-weights coarse scale data; data which is necessary for the large initial changes in disparity required for faster convergence.

In contrast, if we set $\mathcal{O}_{t,q}(\mathbf{s}) = 0$, the GCM converges at a similar rate, but the accuracy becomes worse than all other methods considered, as shown in Table~\ref{table:5FoldValidation} and Figure~\ref{fig:Log_RMSE_plot}.
Figure~\ref{fig:pens_progress} clearly illustrates this in the pens scene; it shows the disparity field rapidly converging to something that roughly resembles the ground truth disparity, but is highly inaccurate around scene-object boundaries. 
In these areas, disparity discontinuities are misshapen and features appear smeared together. 
Figure \ref{fig:pens_3scale} shows that this issue is not resolved over the course of the algorithm.
The poor performance of the GCM when $\mathcal{O}_{t, q}(\mathbf{s})$ is zero is due to the over-weighting of coarse scale data and under-weighting of fine scale data around scene-object boundaries. 
This is especially important since fine scale data is required to properly localise discontinuities in the disparity field.

The dataset used for these tests is composed entirely of synthetic data, and is unaffected by imperfections in the camera calibration and baseline measurements, unlike real-world data. 
When we simulated a small amount of barrel distortion and a small amount of noise in the baseline measurements, we found that the GCM still outperformed the other approaches.

\begin{figure}[!t]
\centering
\includegraphics[width=8.5cm]{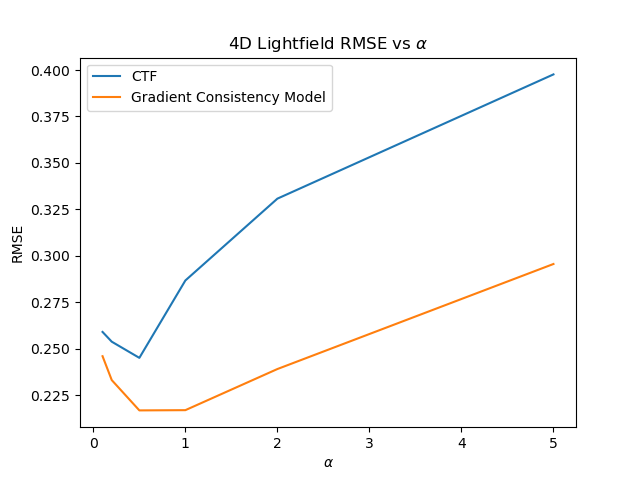}
\caption{Plots of the average final RMSE vs. $\alpha$ across all 24 scenes in the 4D Lightfield Dataset~\cite{Honauer2016}. Here we compare the coarse-to-fine approach (blue) based on \cite{Tran2017Disparity} with the Gradient Consistency Model (orange).}
\label{fig:sensitivity_alpha}
\end{figure}

We also investigate the sensitivity of the GCM to changes in the regularisation parameter $\alpha$. 
In Figure~\ref{fig:sensitivity_alpha} we plot the average final RMSE values of both the coarse-to-fine approach and GCM for different $\alpha$ values.
The $\alpha$ value that performs best for both approaches is 0.5.
However, we can clearly see that the coarse-to-fine scheme is significantly more sensitive to changes in the $\alpha$ than the GCM.

Similarly, we investigate the sensitivity of the GCM to changes in the imaging noise parameter $\epsilon$.
In Table~\ref{table:epsilon_changes} we detail the impact of changing $\epsilon$ across a decade around $\epsilon = 2\times 10^{-4}$.
This has a very small effect on the RMSE value, suggesting that the GCM is substantially insensitive to $\epsilon$ value changes. 

\begin{table}[!t]
\caption{4D Lightfield Dataset Cross-Validation Final RMSE values with different $\epsilon$ values for the GCM.
} \label{table:epsilon_changes}
\centering
\npdecimalsign{.}
\npthousandsep{}
\nprounddigits{5}
\begin{tabular}{c|n{1}{5}}
        \textbf{$\epsilon$} & \textbf{RMSE}  \\ \hline
        $ 2 \times 10^{-4} / \sqrt{10} $ & 0.22236190825 \\ 
    
        $2 \times 10^{-4}$ & 0.224522 \\ 

       $2\sqrt{10} \times 10^{-4}$  & 0.225407  \\ 
\end{tabular}
\npnoround
\end{table}

\subsection{Middlebury 2006 Dataset} \label{ssec:Middlebury_2006}

We also evaluate the performance of the Gradient Consistency Model on the real world multi-view Middlebury 2006 dataset \cite{Hirschmuller2007}. For simplicity of implementation, the half-size images are used rather than the full-size images. We employ the same approach as per Section \ref{ssec:4D_Lightfield} except for several key differences: 
\begin{itemize}
    \item 6 scales are used rather than 3. This is because the Middlebury 2006 dataset has much larger disparities than the 4D Lightfield dataset; the magnitudes of the maximum disparities are generally between 16 and 32.
    \item All 7 views in the dataset are utilised. These are arranged in a co-linear fashion. 
    \item We add a sliding window of scales to the GCM. We only consider three scales simultaneously. Initially, these scales are the three coarsest scales, $q=5$ to $q=3$. We then increment the lowest $q$ and the highest $q$ by one each time we change scales. This approach changes scales at the same points as the coarse-to-fine approach. Considering more scales simultaneously is less practical considering that the GCM approach needs access to all scales in its working set concurrently.
    \item There are 21 scenes in this dataset, so we use 7-fold rather than 6-fold validation to factor out dependence of the performance on the regularisation parameter $\alpha$.  Note, however, that we have already demonstrated the superior insensitivity to $\alpha$ of the GCM in comparison to traditional coarse-to-fine approaches.
\end{itemize}

\begin{figure}[!t]
\centering
\includegraphics[width=8.5cm]{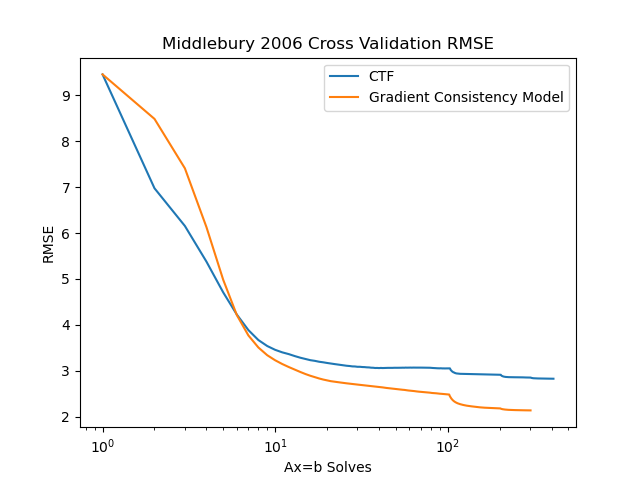}
\caption{Plots of the RMSE vs. the number of $A\mathbf{x} = \mathbf{b}$ solves (log scale) for the average of the 7-fold validation on natural Middlebury 2006 dataset \cite{Hirschmuller2007}. Here we compare the coarse-to-fine approach (blue) based on \cite{Tran2017Disparity} with the Gradient Consistency Model (orange).}
\label{fig:Middlebury_2006_RMSE}
\end{figure}

The average RMSE value for each method after 7-fold validation is plotted against the number of solves in Figure \ref{fig:Middlebury_2006_RMSE}. 
Again, a log scale on the $x$-axis is used to better visualise the early stages of each approach and the final result of each approach is shown in Table \ref{table:Middlebury_2006}. 

\begin{table}[!t]
\caption{Final 7-fold Validation Results on the Middlebury Dataset. 
} \label{table:Middlebury_2006}
\centering
\npdecimalsign{.}
\npthousandsep{}
\nprounddigits{5}
\begin{tabular}{c|n{1}{5}}
        \textbf{Method} & \textbf{RMSE} \\ \hline
        Coarse to Fine & 2.826845383666770 \\ 
    
        Gradient Consistency Model & 2.1100248087789457 \\ 

\end{tabular}
\npnoround
\end{table}

Similar to the results with the 4D Lightfield dataset, the GCM significantly outperforms the coarse-to-fine scheme in terms of both accuracy and rate of convergence. 

\section{Conclusion}\label{sec:Conclusion}


Variational approaches to scene-flow, optical flow and disparity estimation have an elegant theoretical basis. 
However, unfortunately, they often rely on heuristic approaches, such as coarse-to-fine schemes, to address issues with their formulation, including: the limited range over which the linearisation applies; and the inconsistency of the brightness constancy constraints between scales. 
We propose a framework that substantially solves these issues, the Gradient Consistency Model, which is a data-driven, analytically inspired approach which also turns out to be highly robust to changes in the regularisation parameter. 

We have demonstrated the effectiveness of the Gradient Consistency Model in the context of multi-view disparity estimation and shown that it outperforms a coarse-to-fine approach in both accuracy and rate of convergence.

Some or all of the methods proposed in this work could be applied to any variational approach that uses multiple coupled observations of the same scene. 

{\appendices
\section{Derivation of Noise Minimisation Weights}\label{appendix:weights}

In the context of multi-view disparity estimation, the solution to the Euler Lagrange equations, if noise is applied to each $\delta I_{t,q}(\mathbf{s})$ is given by~\eqref{eqn:Noisy_solution}.
To minimise the expected difference between, $\delta {w}^*(\mathbf{s})$ and $\delta {w}^\prime(\mathbf{s})$, we must select weights to minimise,
\begin{multline}\label{eqn:expected_error}
    E[(\delta {w}^\prime(\mathbf{s}) - \delta w^*(\mathbf{s}))^2] = \\
    \frac{ \sum_{t, q} W^2_{t, q}(\mathbf{s}) g^2_{t,q}(\mathbf{s}) E[\mathcal{N}^2_{t, q}(\mathbf{s})] } 
    {(\sum_{t,q} W_{t, q}(\mathbf{s}) g^2_{t,q}(\mathbf{s}))^2},
\end{multline}
assuming that the noise terms are all independent and have a zero mean.
Without any loss of generality, we write $K(\mathbf{s})$ for the sum of the weights $W_{t,q}(\mathbf{s})$ at each location $\mathbf{s}$ — i.e.,  
\begin{equation} \label{eqn:Sum_Weights}
    \sum_{t, q} W_{t, q}(\mathbf{s}) = K(\mathbf{s}).
\end{equation}
Then for any arbitrary set of $K(\mathbf{s})$ values, we can find the individual weights which minimize~\eqref{eqn:expected_error} subject to~\eqref{eqn:Sum_Weights} using the method of Lagrange multipliers.
We write the Lagrangian function of this optimisation problem as,
\begin{equation}
    \mathcal{L} = \lambda K - \lambda \sum_{t,q} W_{t, q}(\mathbf{s}) + E[(\delta {w}^\prime(\mathbf{s}) - \delta w^*(\mathbf{s}))^2].
\end{equation}
The derivative of $\mathcal{L}$ with respect to $W_{k, l}$ is,
\begin{multline}
    \frac{\partial \mathcal{L}}{\partial W_{k, l}(\mathbf{s})} = - \lambda -
    \frac{2g_{k, l}^2(\mathbf{s}) \left[\sum_{t, q} W^2_{t, q}(\mathbf{s}) g^2_{t, q}(\mathbf{s}) E[\mathcal{N}_{t, q}^2(\mathbf{s})] \right]}
{\left[\sum_{t, q} W_{t, q}(\mathbf{s}) g^2_{t, q}(\mathbf{s}) \right]^3} \\ + 
    \frac{2W_{k,l} (\mathbf{s})g_{k, l}^2(\mathbf{s})E[\mathcal{N}_{k, l}^2(\mathbf{s})]\left[\sum_{t, q} W_{t, q}(\mathbf{s}) g^2_{t, q}(\mathbf{s}) \right]}{\left[\sum_{t, q} W_{t, q}(\mathbf{s}) g^2_{t, q}(\mathbf{s}) \right]^3}.
\end{multline}
Solving~$\frac{\partial \mathcal{L}}{\partial W_{k, l}(\mathbf{s})} = 0$  for $W_{k, l}$, we find that
\begin{equation}\label{eqn:Lagrange_Multiplier_Result}
    W_{k, l}(\mathbf{s}) =  \frac{\lambda X_{k, l}(\mathbf{s})}{E[\mathcal{N}^2_{k, l}(\mathbf{s})]} + \frac{Y(\mathbf{s})}{E[\mathcal{N}^2_{k, l}(\mathbf{s})]},
\end{equation}
where
\begin{equation}
    X_{k, l}(\mathbf{s}) = \frac{(\sum_{t, q} W_{t, q}(\mathbf{s}) g^2_{t, q}(\mathbf{s}))^2}
    {2 g_{k, l}^2(\mathbf{s}) },
\end{equation}
and
\begin{equation}
    Y(\mathbf{s}) = \frac{\sum_{t, q} W^2_{t,q}(\mathbf{s}) g^2_{t, q}(\mathbf{s}) E[\mathcal{N}_{t, q}^2(\mathbf{s})]}{\sum_{t, q} W_{t, q}(\mathbf{s}) g^2_{t, q}(\mathbf{s}) }.
\end{equation}
The constraint~\eqref{eqn:Sum_Weights} can be applied to~\eqref{eqn:Lagrange_Multiplier_Result} yielding,
\begin{equation}\label{eqn:applied_constraint_weights}
    \lambda \sum_{m, n} \frac{X_{m, n}(\mathbf{s})}{E[\mathcal{N}^2_{m, n}(\mathbf{s})]} + \sum_{m, n} \frac{Y(\mathbf{s})}{E[\mathcal{N}^2_{m, n}(\mathbf{s})]}  = K(\mathbf{s}).
\end{equation}
We can solve~\eqref{eqn:applied_constraint_weights} for $\lambda$, 
\begin{equation}\label{eqn:lambda_value}
    \lambda = \frac{K(\mathbf{s}) - \sum_{m, n}Y(\mathbf{s})/E[\mathcal{N}^2_{m, n}(\mathbf{s})]}
    {\sum_{m,n} X_{m, n}(\mathbf{s})/E[\mathcal{N}^2_{m, n}(\mathbf{s})]}.
\end{equation}
Substituting~\eqref{eqn:lambda_value} into~\eqref{eqn:Lagrange_Multiplier_Result}, we find that
\begin{multline}
     W_{k, l}(\mathbf{s}) = \frac{Y(\mathbf{s})}{E[\mathcal{N}^2_{k, l}(\mathbf{s})]} + \\
     \frac{X_{k, l}(\mathbf{s}) }{E[\mathcal{N}^2_{k, l}(\mathbf{s})]}\frac{K(\mathbf{s}) - \sum_{m, n}Y(\mathbf{s})/E[\mathcal{N}^2_{m, n}(\mathbf{s})]}
    {\sum_{m,n} X_{m, n}(\mathbf{s})/E[\mathcal{N}^2_{m, n}(\mathbf{s})]}.
\end{multline}
In the end, $K(\mathbf{s})$ itself is an arbitrary choice that does not affect the optimality of the resulting individual weights, so long as the useless choice of $K(\mathbf{s}) = 0$ is avoided. 
Therefore, in practice we use,
\begin{equation}
    W_{t, q}(\mathbf{s}) = \frac{Z}{E[\mathcal{N}^2_{t, q}(\mathbf{s})]},
\end{equation}
where $Z$ is an arbitrary positive constant. 

\section{Derivation of Scale Inconsistency}\label{appendix:Spatial}

The purpose of this appendix is to derive an upper bound for the $\mathcal{O}^2_{p,q}(\mathbf{s}) = (J_A(\mathbf{s}) - J_B(\mathbf{s}))^2$ term that represents the deviation of the data term from the low-pass filtered linearised brightness constancy constraints that are consistent with the ideal brightness constancy constraint.

In a multi-view disparity estimation context, we write $J_A(\mathbf{s})$ as,
\begin{equation}
    J_A(\mathbf{s}) = \iint_\Omega G_{\sigma_q}(\boldsymbol{\tau}) g_{t, 0}(\mathbf{s} - \boldsymbol{\tau}) \widetilde{ \delta w}(\mathbf{s} - \boldsymbol{\tau}) d \boldsymbol{\tau} + \delta I_{t, q}(\mathbf{s}),
\end{equation}
and $J_B(\mathbf{s})$ as,
\begin{equation}
    J_B(\mathbf{s}) = \widetilde{ \delta w}(\mathbf{s}) \iint_\Omega G_{\sigma_q}(\boldsymbol{\tau}) g_{t,0}(\mathbf{s} - \boldsymbol{\tau})  d \boldsymbol{\tau} + \delta I_{t,q}(\mathbf{s}).
\end{equation}
If $\widetilde{ \delta w}(\mathbf{s})$ is a constant, $J_A(\mathbf{s}) = J_B(\mathbf{s})$. 
Hence, $\mathcal{O}^2_{t,q}(\mathbf{s}) = 0$.

At the opposite extreme, suppose the displacement field is highly correlated with the gradient field. 
In particular, we consider the specific case where, 
\begin{equation}\label{eqn:disparity2grad_proportional}
    \widetilde{ \delta w}(\mathbf{s}) = \gamma \cdot g_{t, 0}(\mathbf{s}).
\end{equation}
Such a relationship is unreasonable to expect in general, but may apply locally.
In this case, we can write $\mathcal{O}^2_{t,q}(\mathbf{s})$ as,
\begin{multline}
    \mathcal{O}^2_{t, q}(\mathbf{s}) = 
    \left(  \vphantom{\iint_\Omega}
        \smash{\gamma \cdot \overbrace{\iint_\Omega G_{\sigma_q}(\boldsymbol{\tau}) g_{t, 0}^2(\mathbf{s} - \boldsymbol{\tau}) d \boldsymbol{\tau}}^{J^\prime_A(\mathbf{s})}} \right. 
        \vphantom{\overbrace{\iint_\Omega}}
        \\
        \left. - \gamma \cdot \vphantom{\iint_\Omega}
        \smash{\underbrace{g_{t,0}(\mathbf{s}) \iint_\Omega G_{\sigma_q}(\boldsymbol{\tau}) g_{t,0}(\mathbf{s} - \boldsymbol{\tau})  d \boldsymbol{\tau}}_{J_B^\prime(\mathbf{s})}} \hphantom{i}
    \right)^2 \vphantom{\underbrace{\iint_\Omega}_{J^\prime_B(\mathbf{s})}}.
\end{multline}
In this extreme case, the $(J^{\prime}_A)^2(\mathbf{s})$ can be written as,
\begin{multline} \label{eq:spatial_grad_consistency_special_case}
    (J^\prime_A)^2(\mathbf{s}) = 
    \left(
        \iint_\Omega G_{\sigma_q}(\boldsymbol{\tau}) g_{t,0}^2(\mathbf{s} - \boldsymbol{\tau}) d \boldsymbol{\tau} \right)\\
        \cdot \left(
        \iint_\Omega G_{\sigma_q}(\boldsymbol{\tau}) \widetilde{ \delta w}^2(\mathbf{s} - \boldsymbol{\tau}) d \boldsymbol{\tau} \right),
\end{multline}
which can be readily estimated. 
Moreover under these extreme conditions, it turns out that $(J^{\prime}_A)^2(\mathbf{s})$ dominates $\mathcal{O}^2_{t,q}(\mathbf{s})$, ie
\begin{equation}
    \mathcal{O}^2_{t,q}(\mathbf{s}) \approx (J^{\prime}_A)^2(\mathbf{s}).
\end{equation}
To see this, observe that $J^\prime_A(\mathbf{s})$ is a low-pass filtered non-negative function; whereas, $J^\prime_B(\mathbf{s})$ is the product of a high pass and a low-pass signal. 
We could construct various elaborate arguments to explain why $J^\prime_B(\mathbf{s})$ becomes vanishingly small in comparison to $J^\prime_A(\mathbf{s})$ as $\sigma_q$ becomes large. However, to keep things simple, we choose instead to provide a plot of the ratio between the expected power of $J^\prime_A(\mathbf{s})$ and $J^\prime_B(\mathbf{s})$ vs $\sigma_q$ - Figure~\ref{fig:J_A_vs_J_B}. 

\begin{figure}[!t]
\centering
\includegraphics[width=8.5cm]{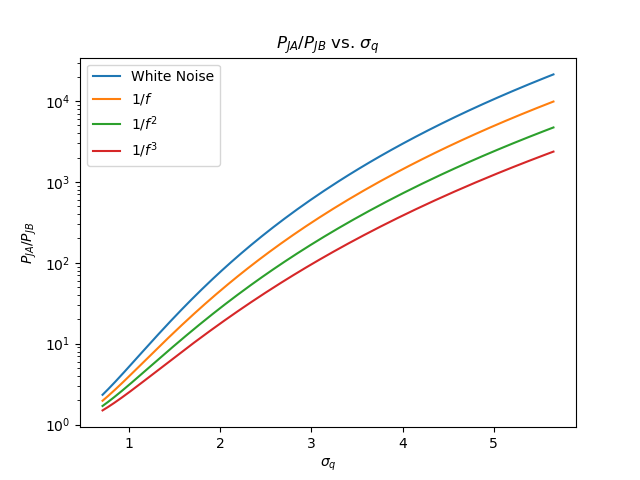}
\caption{Plots of the ratio between the expected power of $J^\prime_A(\mathbf{s})$ ($P_{JA}$) and $J^\prime_B(\mathbf{s})$ ($P_{JB}$) vs $\sigma_q$ with various models of the image power spectrum. These are: white noise; $1/f$; $1/f^2$; and $1/f^3$. A log scale is used on the $y$-axis. For $g_{t, 0}$, we used a Derivative of Gaussians filter with $\sigma_0 = 1/\sqrt{2}$ in the $s_1$ direction applied to the image power spectrum. We also limit the frequency domain to $[-\pi, \pi]$.
}
\label{fig:J_A_vs_J_B}
\end{figure}

\section{Derivation of Image Acquisition Noise Term}\label{appendix:ImagingNoise}
We assume that image acquisition noise, $\mathcal{A}(\mathbf{s})$ is additive Gaussian white noise applied to the linearised brightness constancy constraint~\eqref{eqn:brightConstLin}. After filtering by a Gaussian filter $G_{\sigma_q}(\mathbf{s})$, this noise term can be written as,
\begin{equation}
    \varepsilon(\mathbf{s}) = (G_{\sigma_q} * \mathcal{A})(\mathbf{s}).
\end{equation}
The power spectrum of $\varepsilon(\mathbf{s})$ can be written as,
\begin{equation}\label{eqn:img_noise_pwr_spectrum}
    |\hat{\varepsilon}(\boldsymbol{\omega})|^2 = |\hat{G}_{\sigma_q}(\boldsymbol{\omega})|^2 |\hat{\mathcal{A}}(\boldsymbol{\omega})|^2.
\end{equation}
Since $\mathcal{A}(\mathbf{s})$ is white noise, its power spectrum can be written $|\hat{\mathcal{A}}(\boldsymbol{\omega})|^2 = \epsilon^2$. 
Accordingly, we can simplify~\eqref{eqn:img_noise_pwr_spectrum},
\begin{equation}
    |\hat{\varepsilon}(\boldsymbol{\omega})|^2 = \epsilon^2 |\hat{G}_{\sigma_q}(\boldsymbol{\omega})|^2.
\end{equation}
Accordingly, we can calculate the total power by integrating $|\hat{\varepsilon}(\boldsymbol{\omega})|^2$ from $-\infty$ to $\infty$ in both directions. 
Specifically,
\begin{equation}
    \mathcal{I} = \frac{\epsilon^2}{4\pi^2} \int^\infty_{-\infty} e^{-\sigma_q^2 \omega_1^2} d\omega_1 \int^\infty_{-\infty} e^{-\sigma_q^2 \omega_2^2} d\omega_2.
\end{equation}
Evaluating these integrals yields,
\begin{equation}
    \mathcal{I} = \frac{\epsilon^2 \pi}{16\pi^2\sigma^2} | \text{erf}( \sigma_q \omega_1)|^\infty_{-\infty} \times |\text{erf}( \sigma_q \omega_2)|^\infty_{-\infty}\\ 
= \frac{\epsilon^2}{4 \pi \sigma_q^2}.
\end{equation}

}

\bibliographystyle{IEEEtran}
\bibliography{bibliography}

\end{document}